\documentclass[12pt,preprint]{aastex}
\usepackage{}
\usepackage{amssymb}
\usepackage{amsmath}
\usepackage{graphicx}
\usepackage{natbib}
\usepackage{tabularx}
\usepackage{epsfig}
\newcommand{\AV}{A_V}
\newcommand{\Cext}{C_{\rm ext}}

\newcommand{\tausil}{\Delta\tau_{9.7}}
\newcommand{\Cextsil}{C_{\rm ext}(9.7\,\mu {\rm m})}
\newcommand{\CextVsil}{C_{\rm ext}^{\rm sil}(V)}
\newcommand{\CextVcarb}{C_{\rm ext}^{\rm carb}(V)}
\newcommand{\aeq}{a_{\rm eq}}
\newcommand       \tauahc       {\Delta\tau_{3.4}}
\newcommand       \sgrA         {{\rm Sgr}\,{\rm A}^{\ast}}
\newcommand	\beq	{\begin{equation}}	
\newcommand	\eeq	{\end{equation}}	
\newcommand       \Angstrom     {\,{\rm \AA}}

\newcommand       \cm           {\,{\rm cm}}

\newcommand       \g            {\,{\rm g}}

\newcommand       \NH           {N_{\rm H}}
\newcommand       \simlt        {\lesssim}
\newcommand       \simgt        {\gtrsim}
\newcommand       \gtsim        {\gtrsim}

\newcommand       \mum          {\,{\rm \mu m}}
\newcommand       \ppm          {\,{\rm ppm}}

\newcommand       \simali       {\sim\,}
\newcommand       \magni        {\,{\rm mag}}

\newcommand       \Msil           {M_{\rm sil}}
\newcommand       \Mcarb          {M_{\rm carb}}
\newcommand       \rhosil        {\rho_{\rm sil}}
\newcommand       \rhocarb        {\rho_{\rm carb}}

\newcommand	  \csun         {\left[{\rm C/H}\right]_{\odot}}

\newcommand	  \fesun        {\left[{\rm Fe/H}\right]_{\odot}}
\newcommand	  \mgsun        {\left[{\rm Mg/H}\right]_{\odot}}
\newcommand	  \sisun        {\left[{\rm Si/H}\right]_{\odot}}
\newcommand	  \cism         {\left[{\rm C/H}\right]_{\rm ISM}}

\newcommand	  \cdust        {\left[{\rm C/H}\right]_{\rm dust}}

\newcommand	  \fedust       {\left[{\rm Fe/H}\right]_{\rm dust}}
\newcommand	  \mgdust       {\left[{\rm Mg/H}\right]_{\rm dust}}
\newcommand	  \sidust       {\left[{\rm Si/H}\right]_{\rm dust}}

\newcommand	  \cgas        {\left[{\rm C/H}\right]_{\rm gas}}

\newcommand	  \muc         {\mu_{\rm C}}

\newcommand	  \musi        {\mu_{\rm Si}}
\newcommand	  \mufe        {\mu_{\rm Fe}}
\newcommand	  \mumg        {\mu_{\rm Mg}}

\countdef\decade=200
\advance\decade by \year
\countdef\hours=201
\advance\hours by \time
\divide\hours by 60
\countdef\mins=202
\advance\mins by \hours
\multiply\mins by 60
\multiply\hours by 100
\countdef\miltime=203
\advance\miltime by \hours
\advance\miltime by \time
\advance\miltime by -\mins
\def\today{\number\decade.\number\month.\number\day.\number\miltime}
\shorttitle{Dust in AGNs}
\begin{document}
\title{On the Optical-to-Silicate Extinction Ratio
          as a Probe of the Dust Size
          in Active Galactic Nuclei}
\author{
Zhenzhen~Shao\altaffilmark{1,2},
B.W.~Jiang\altaffilmark{1}, and
Aigen~Li\altaffilmark{2}
}
\altaffiltext{1}{Department of Astronomy,
                       Beijing Normal University,
                       Beijing 100875, China;
                       {\sf zhenzhenshao@mail.bnu.edu.cn,
                              bjiang@bnu.edu.cn}
                       }
\altaffiltext{2}{Department of Physics and Astronomy,
                       University of Missouri,
                       Columbia, MO 65211, USA;
                       {\sf lia@missouri.edu}
                       }

\begin{abstract}
Dust plays a central role in the unification theory of
active galactic nuclei (AGNs). Whether the dust that
forms the torus around an AGN is tenth micron-sized
like interstellar grains or much larger has
a profound impact on correcting for the obscuration
of the dust torus to recover the intrinsic spectrum
and luminosity of the AGN. Here we show that
the ratio of the optical extinction in the visual band
($\AV$) to the optical depth of the 9.7$\mum$ silicate
absorption feature ($\tausil$) could potentially be
an effective probe of the dust size.
The anomalously lower ratio of
$\AV/\tausil\approx5.5$ of AGNs
compared to that of the Galactic
diffuse interstellar medium of
$\AV/\tausil\approx18$
reveals that the dust in AGN torus could be
substantially larger than the interstellar
grains of the Milky Way and of
the Small Magellanic Cloud,
and therefore, one might expect a flat
extinction curve for AGNs.
\end{abstract}



\section{Introduction\label{intro}}
Silicate dust is ubiquitously seen in a wide variety of
astrophysical environments through the absorption
or emission spectral features arising from
the Si--O and O--Si--O vibrational modes
(see Henning 2010).
Interstellar silicate dust is of particular interest
to astronomers, not only because it is a major dust
species in the interstellar medium (ISM) and accounts
for a substantial fraction of the observed optical extinction,
but also because it is the raw material for building
the obscuring torus around active galactic nuclei
(AGNs; see Siebenmorgen \& Kr\"ugel 2007)
and the planet-forming disks around young stars
(e.g., see Kr\"ugel \& Siebenmorgen 1994).
Since most of our knowledge about interstellar dust
is inferred from its interaction with the electromagnetic
radiation, the size of the dust is a key parameter in
characterizing the properties of the dust (see Li 2009).
While the characteristic size of interstellar silicate dust
is around 0.1$\mum$,\footnote{%
  This characteristic size of $\simali$0.1$\mum$
   is inferred from the interstellar visual extinction.
   According to light scattering theory,
   dust grains of sizes comparable to
   the wavelength ($\lambda$) of the incident light
   absorb and scatter photons most effectively
   (i.e., $2\pi a/\lambda\sim1$, where $a$
   is the spherical radius of the grain; see Li 2009).
   With $\lambda_V\approx5500\Angstrom$
   for the visual band, one obtains
   $a\sim\lambda/2\pi\sim0.1\mum$.
   }
the sizes of the silicate dust grains
in the torus around AGNs and
in protoplanetary disks around young stars
are less clear.
The predominant presence of larger grains
in AGN torus or protoplanetary disks
than their interstellar counterpart
has important astrophysical implications.
For protoplanetary disks, large grains
imply the occurrence of grain growth,
the first step toward planet formation.
For AGN torus, larger grains indicate
a smaller amount of extinction
(on a per unit dust mass basis)
and a flatter extinction curve
in the ultraviolet (UV) and
optical wavelength range.

In the diffuse ISM,
the Si--O and O--Si--O
stretching features occur at
$\simali$9.7 and $\simali$18$\mum$,
respectively, and are seen in absorption
(e.g., see Siebenmorgen et al.\ 2014).
These features are smooth
and lack fine structures,
indicating a predominantly
amorphous composition
(Li \& Draine 2001,
Kemper et al.\ 2004,
Li et al.\ 2007).
In protoplanetary disks,
the 9.7$\mum$ Si--O feature,
usually seen in emission, is often
broader than the interstellar
absorption feature. This is commonly
interpreted as grain growth.
Another piece of observational evidence
for grain growth in protoplanetary disks
comes from the far infrared (IR) to
submillimeter opacity spectral index
which is often flatter than its
interstellar counterpart
(see Draine 2006).

For AGNs, the situation is more complicated.
The unification theory of AGNs assumes that
all AGNs are surrounded by an anisotropic
dusty torus and are essentially the same objects
but viewed from different angles
(Antonucci 1993, Urry \& Padovani 1995).
According to this theory, type 1 AGNs
are viewed  face-on and unobscured,
while type 2 AGNs are viewed edge-on
and the central regions, including the black hole,
accretion disk and broad line region, are obscured.
For type 1 AGNs, the Si--O stretching feature
is mostly seen in emission
(Hao et al.\ 2005, Siebenmorgen et al.\ 2005,
Sturm et al.\ 2005), as expected from
the unification theory. The peak wavelength
of this feature often appreciably shifts
from the canonical wavelength of
$\simali$9.7$\mum$ to longer wavelengths
beyond $\simali$10$\mum$.
While Li et al.\ (2008) and Smith et al.\ (2010)
attributed this longward wavelength shift
of the ``9.7$\mum$'' Si--O emission feature
to $\mu$m-sized silicate grains,
Nikutta et al.\ (2009) argued that,
in the framework of a clumpy dust torus,
a mixture of sub-$\mu$m-sized
interstellar silicate and graphite grains
can explain the observed longward wavelength shift
of the silicate emission feature.
They suggested that the longward wavelength shift
could just be caused by radiation transfer effects.
Similarly, Siebenmorgen et al.\ (2005, 2015)
also found that their self-consistent
two-phase AGN torus model could closely
explain the observed wavelength shift
of the ``9.7$\mum$'' Si--O feature entirely
in terms of radiative transfer effects.

Whether the dust in the AGN torus is
sub-$\mu$m-sized like its precursor
(i.e., interstellar grains) or much larger
like that in protoplanetary disks
has important impact on correcting for
the dust extinction to recover the intrinsic
spectra and luminosity of AGNs.
If the AGN dust is interstellar-sized
or smaller, one would expect an extinction curve
like that of the Milky Way
or the Small Magellanic Cloud (SMC)
of which the extinction $A_\lambda$
steeply rises with the increase
of $\lambda^{-1}$, the inverse wavelength.
In contrast, if the AGN dust is $\mu$m-sized,
one would then expect a flat or gray extinction curve
which exhibits little variation with wavelength
(e.g., see Figure~7 in Xie et al.\ 2017).
Numerous efforts have been made to deduce
the AGN torus extinction curve but lead to
contradicting results (see Li 2007 for a review).
In the literature, the AGN extinction curves are mainly
inferred from (1) composite quasar spectra, and
(2) individual reddened AGNs. The former often reveals
a ``gray'' extinction, implying that the size distribution
of the dust in the AGN circumnuclear environments is
skewed toward substantially large grains
(e.g., see Gaskell et al.\ 2004, Czerny et al.\ 2004,
Gaskell \& Benker 2007).
The latter often suggests a steeply-rising SMC-like extinction,
indicating a preponderance of small grains near the nucleus
(e.g., see Richard et al.\ 2003,
Hopkins et al.\ 2004, Glikman et al.\ 2012).
However, Siebenmorgen et al.\ (2015)
argued that the AGN extinction curve
is not necessarily a telltale of the dust size
since there is no direct one-to-one link
between the observed (or apparent) extinction
curve of AGNs and the wavelength-dependence of
the total dust cross sections. The latter is required
to determine the dust size.

\begin{figure}[h]
\centering
\includegraphics[height=8cm,width=15cm]{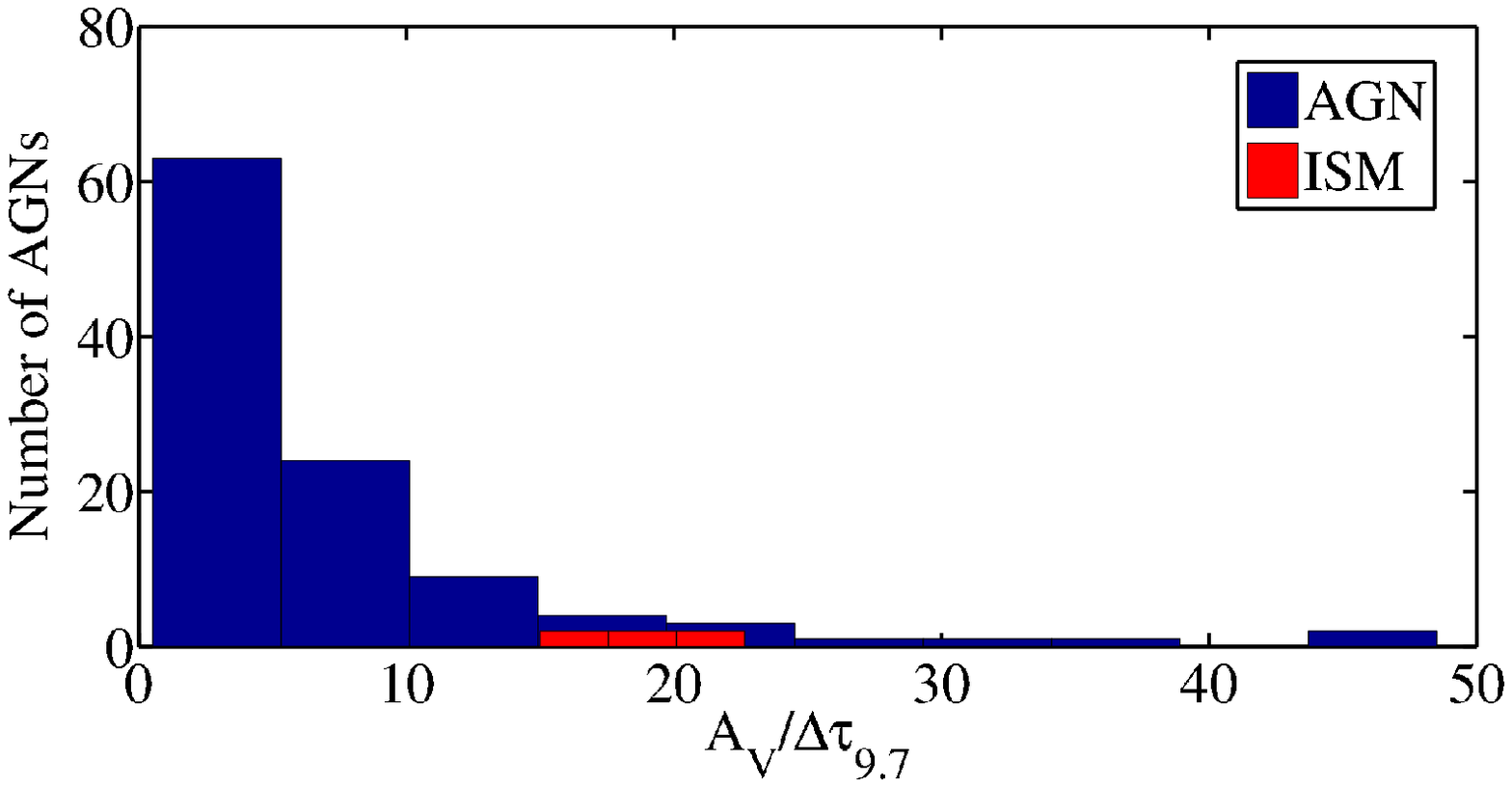}
\caption{\footnotesize
               \label{fig:histogram}
               Histograms of the optical-to-silicate
               extinction ratios for 110 type 2 AGNs (blue)
               measured by Lyu et al.\ (2014)
               and for six lines of sight toward
               the solar neighbourhood diffuse ISM (red)
               measured by Roche \& Aitken (1984).
               }
\end{figure}

In this work we propose
an alternative diagnosis of the dust size
in AGN torus, based on $A_V/\Delta\tau_{9.7}$,
the ratio of the visual extinction $A_V$
to the 9.7$\mum$ silicate absorption
optical depth $\Delta\tau_{9.7}$.
Roche \& Aitken (1984) determined
$A_V/\Delta\tau_{9.7}\approx18$
for the solar neighbourhood diffuse ISM
from six bright WC8 or WC9 Wolf-Rayet stars
which suffer a visual extinction in the range of
$4<\AV<17\magni$. The extinction to these stars
was found to be dominantly interstellar origin
with little extinction from their circumstellar shells.
In contrast, Lyu et al.\ (2014) derived
$A_V/\Delta\tau_{9.7}\approx5.5$
for 110 type 2 AGNs from the Balmer decrement
measured with the Sloan Digital Sky Survey (SDSS)
and from the 9.7$\mum$ absorption spectra
measured with the {\it Spitzer}/Infrared Spectrograph (IRS).
As shown in Figure~\ref{fig:histogram},
although the $\AV/\tausil$ ratio of
the AGN sample of Lyu et al.\ (2014)
exhibits a much larger scatter
than that of the ISM sample of
Roche \& Aitken (1984),
it is apparent that the majority
of the AGN sample has
an $\AV/\tausil$ ratio
much smaller than that of the diffuse ISM.
Here we will show that,
compared to the local ISM's mean ratio
of $A_V/\Delta\tau_{9.7}\approx18$
(Roche \& Aitken 1984),
the anomalously low ratio of
$A_V/\Delta\tau_{9.7}\approx5.5$
derived for 110 type 2 AGNs (Lyu et al.\ 2014)
implies that the dust in AGN torus is
substantially larger than the interstellar
grains of the Milky Way and of the SMC.
This is independent of the exact dust
shape and composition, as demonstrated
in \S\ref{sec:mod}. We will discuss in
\S\ref{sec:discussion} the role of carbon dust
and the resulting flat extinction curve for AGNs
and summarize the major results
in \S\ref{sec:summary}.
We will focus on type 2 AGNs
as in type 1 AGNs and protoplanetary disks
the 9.7$\mum$ silicate feature is often
seen in emission.

\section{$A_V/\tausil$ as a Probe
            of Dust Size\label{sec:mod}}
\begin{figure}[h]
\centering
\includegraphics[height=8cm,width=15cm]{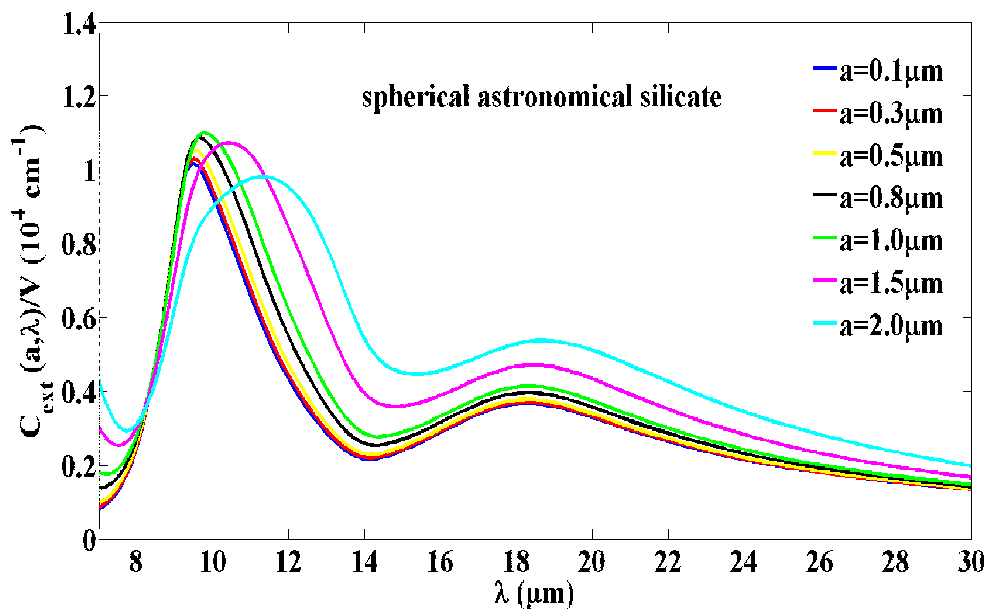}
\caption{\footnotesize
               \label{fig:Cextsil}
               Extinction cross sections (per unit volume)
               of spherical ``astronomical silicate'' grains
               of various radii.
               }
\end{figure}

We first consider spherical grains of radii $a$
and of ``astronomical silicate'' composition
(Draine \& Lee 1984).
We use Mie theory to calculate its
extinction cross section $\Cext(a,\lambda)$
at wavelength $\lambda$. The ratio of $A_V$
to the 9.7$\mum$ silicate absorption depth
$\Delta\tau_{9.7}$ is
\begin{equation}\label{eq:AV2tausil}
\frac{\AV}{\Delta\tau_{9.7}}=\frac{1.086\,\Cext(V)}
        {\Delta\Cext(9.7\mu {\rm m})} ~~,
\end{equation}
where $\Cext(V)$ is the extinction cross section
at the visual band (i.e., $\lambda_V=5500\Angstrom$),
and $\Delta\Cextsil$ is the continuum-subtracted
excess cross section of the 9.7$\mum$ silicate feature.
In Figure~\ref{fig:Cextsil} we show the extinction cross
sections of spherical ``astronomical silicate'' of
various sizes in the wavelength range of
the 9.7 and 18$\mum$ silicate features.
On a per unit volume basis, we see that the extinction
profiles are almost identical for $a\simlt0.5\mum$,
implying that grains of $a\simlt0.5\mum$ are
in the Rayleigh regime (i.e., $2\pi a/\lambda \ll 1$).
For grains of $a\simgt0.8\mum$, the 9.7$\mum$
Si--O absorption feature becomes appreciably broader
and its peak wavelength ($\lambda_{\rm peak}$)
shifts to longer wavelength, e.g., $\lambda_{\rm peak}$
red-ward shifts from $\simali$9.5$\mum$ for $a=0.1\mum$
to $\simali$11.4$\mum$ for $a=2.0\mum$,
while the full-width-half-maximum (FWHM)
of this feature increases from $\simali$2.59$\mum$
to $\simali$5.50$\mum$. 
We note that the red-ward wavelength shift
of $\lambda_{\rm peak}$ and the FWHM-broadening
due to grain-size increase naturally explains
what have been observed in AGNs,
e.g., Hao et al.\ (2005) found that the ``9.7$\mum$''
silicate emission feature of the luminous quasar 3C\,273
peaks at $\simali$11$\mum$; Sturm et al.\ (2005)
and Smith et al.\ (2010) detected similar phenomena
in the low luminosity AGN NGC 3998 and the nucleus
of M81, respectively.

\begin{figure}[h]
\centering
\includegraphics[width=.8\textwidth]{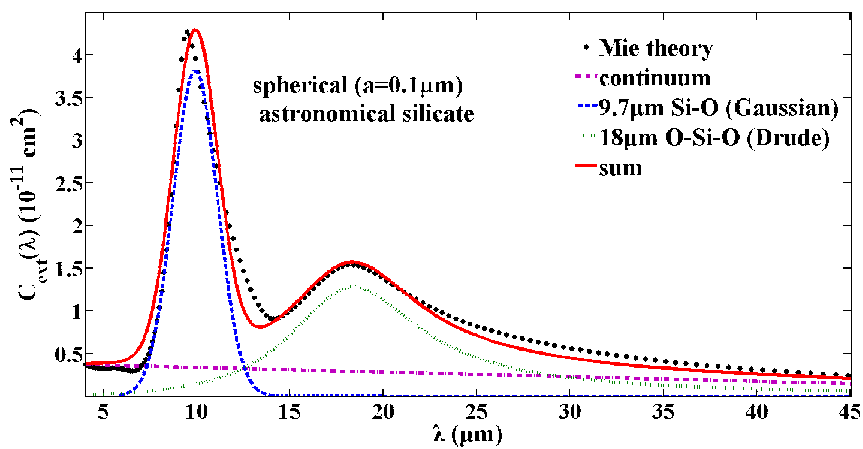}
\caption{\footnotesize
               \label{fig:continuum}
               Fitting the extinction profile of spherical
               ``astronomical silicate'' of radii $a=0.1\mum$
               calculated from Mie theory
               with a Gaussian profile for
               the 9.7$\mum$ Si--O feature
               and a Drude profile for
               the 18$\mum$ O--Si--O feature
               together with an underlying linear continuum.
               }
\end{figure}

\begin{figure}[h]
\centering
\includegraphics[width=.8\textwidth]{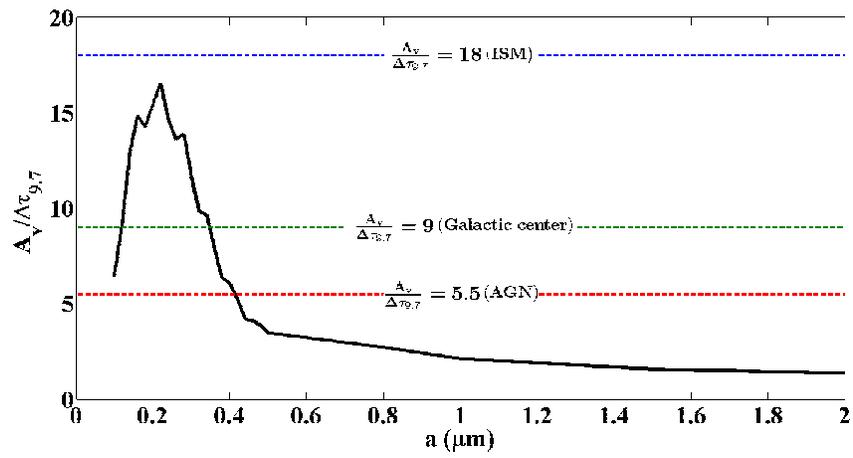}
\caption{\footnotesize
              \label{fig:AV2Sil}
              $\AV/\tausil$ of spherical ``astronomical silicate''
              as a function of dust size $a$ (black solid line).
              Also shown are the observed $\AV/\tausil$ ratios
              for AGNs (red dashed line),
              the Galactic center (green dashed line), and
              the local diffuse ISM (blue dashed line).
              }
\end{figure}

To subtract the continuum underneath
the 9.7 and 18$\mum$ silicate features,
we fit these two features with parameterized functions
together with an underlying linear continuum.
As demonstrated in Figure~\ref{fig:continuum},
it turns out that a Gaussian profile better fits
the 9.7$\mum$ feature while a Drude profile
fits the 18$\mum$ feature better.
In Figure~\ref{fig:AV2Sil} we show the ratio
of the visual extinction to the continuum-subtracted
9.7$\mum$ optical depth as a function of grain size,
calculated from spherical ``astronomical silicate'' grains.
It is seen that $A_V/\tausil$ first rapidly increases with $a$
before it reaches a peak value of $\simali$16.5
at $a\approx0.22\mum$, then it quickly decreases
with $a$ till $a\simlt0.5\mum$, and finally,
at $a>0.5\mum$, the decrease of $A_V/\tausil$
with $a$ becomes more gradual.
For the AGN ratio of $\AV/\tausil\approx5.5$
(Lyu et al.\ 2014), from Figure~\ref{fig:AV2Sil}
one derives the dust size to be
$\simali$0.42$\mum$.\footnote{%
   Although it appears that one could also
   achieve $\AV/\tausil\approx5.5$ with
   $a\approx0.09\mum$, we note that it is unlikely
   for such small grains to survive in the hostile
   circumnuclear environments around AGNs
   (e.g., see Siebenmorgen et al.\ 2004).
   The spectroscopic studies of the 9.7$\mum$
    silicate feature of AGNs all point to dust much
    larger than $\simali$0.1$\mum$
    (e.g., see Li et al.\ 2008, K\"{o}hler \& Li 2010,
     Smith et al.\ 2010, Xie et al.\ 2017).
    }
We note that, as will be shown in \S\ref{sec:discussion},
this size of $a\approx0.42\mum$ is a lower limit
since silicate dust is not the sole contributor
to the optical extinction.

\begin{figure}[ht]
\centering
\includegraphics[width=.8\textwidth]{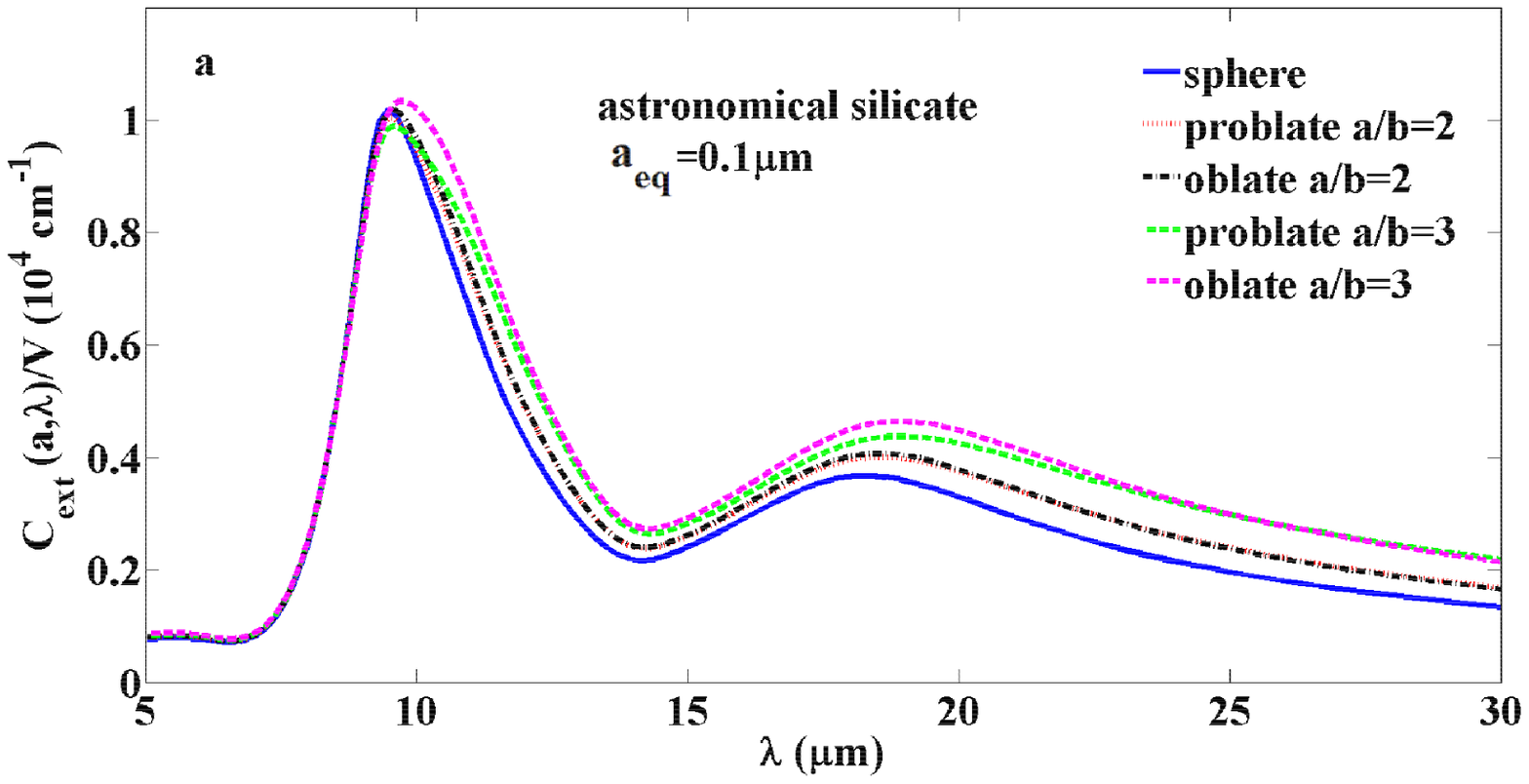}
\includegraphics[width=.8\textwidth]{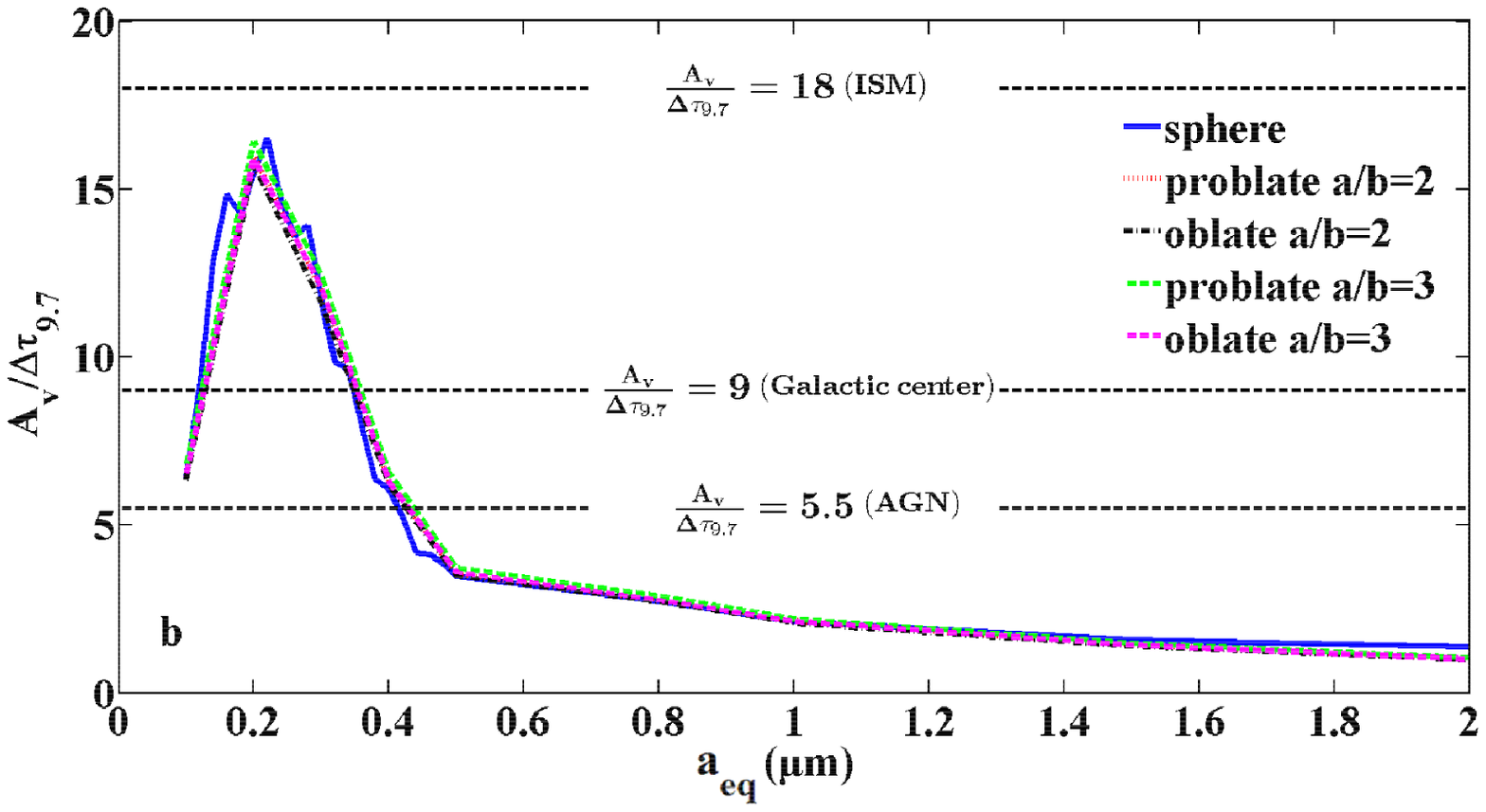}
\caption{\footnotesize
               \label{fig:shape}
               Upper panel (a): extinction cross sections
               (on a per unit volume basis) of spherical
               and spheroidal ``astronomical silicate''
               grains of $\aeq=0.1\mum$.
               Lower panel (b): Same as Figure~\ref{fig:AV2Sil}
               but for spheroidal grains.
               }
\end{figure}

So far, we have assumed the dust to be spherical.
It is well recognized that interstellar grains,
the precursor of the dust grains in AGN torus,
are nonspherical as revealed by the detection of
interstellar polarization
(e.g., see Siebenmorgen et al.\ 2014).
%
To examine the effects of grain shape on $\AV/\tausil$,
we consider spheroidal grains.
Let $a$ and $b$ respectively be
the semiaxis along and perpendicular
to the symmetry axis of a spheroidal grain.
Let $\aeq\equiv\left(a^2b\right)^{1/3}$
be the radius of the equal-volume sphere.
We consider prolates of $a/b=2,\,3$ and
oblates of $b/a=2,\,3$.
In Figure~\ref{fig:shape}a
we show the extinction cross sections
calculated from the Rayleigh scattering
approximation (Bohren \& Huffman 1983)
for these spheroids of $\aeq=0.1\mum$
and of ``astronomical silicate'' composition.
It is seen in Figure~\ref{fig:shape}a that
the extinction profiles of spheroidal grains
do not differ substantially from that of
spherical grains, except for more elongated
grains, the red wing of the 9.7$\mum$ feature
becomes broader and the overall extinction
level of the 18$\mum$ feature becomes higher.
We have also explored the variation of
$\AV/\tausil$ with $\aeq$
for spheroidal grains.\footnote{%
   At the wavelength range of
   the 9.7 and 18$\mum$ silicate features,
   the Rayleigh scattering approximation
   is valid for $\aeq\simlt0.5\mum$
   (see Figure~\ref{fig:Cextsil}).
   For the visual band, the Rayleigh approximation
   is not valid for $a\gtsim0.05\mum$.
   We therefore calculate $\AV$ with Mie theory
   from their equal-volume spheres.
   }
As shown in Figure~\ref{fig:shape}b,
the results are closely similar to
that of spherical grains:
$\AV/\tausil$ peaks at $\aeq\approx0.2\mum$
and for the AGN ratio of $\AV/\tausil\approx5.5$
one derives $\aeq\approx0.4\mum$.
This demonstrates that grain shapes do
not appreciably affect $\AV/\tausil$.

\begin{figure}[h]
\centering
\includegraphics[width=.8\textwidth]{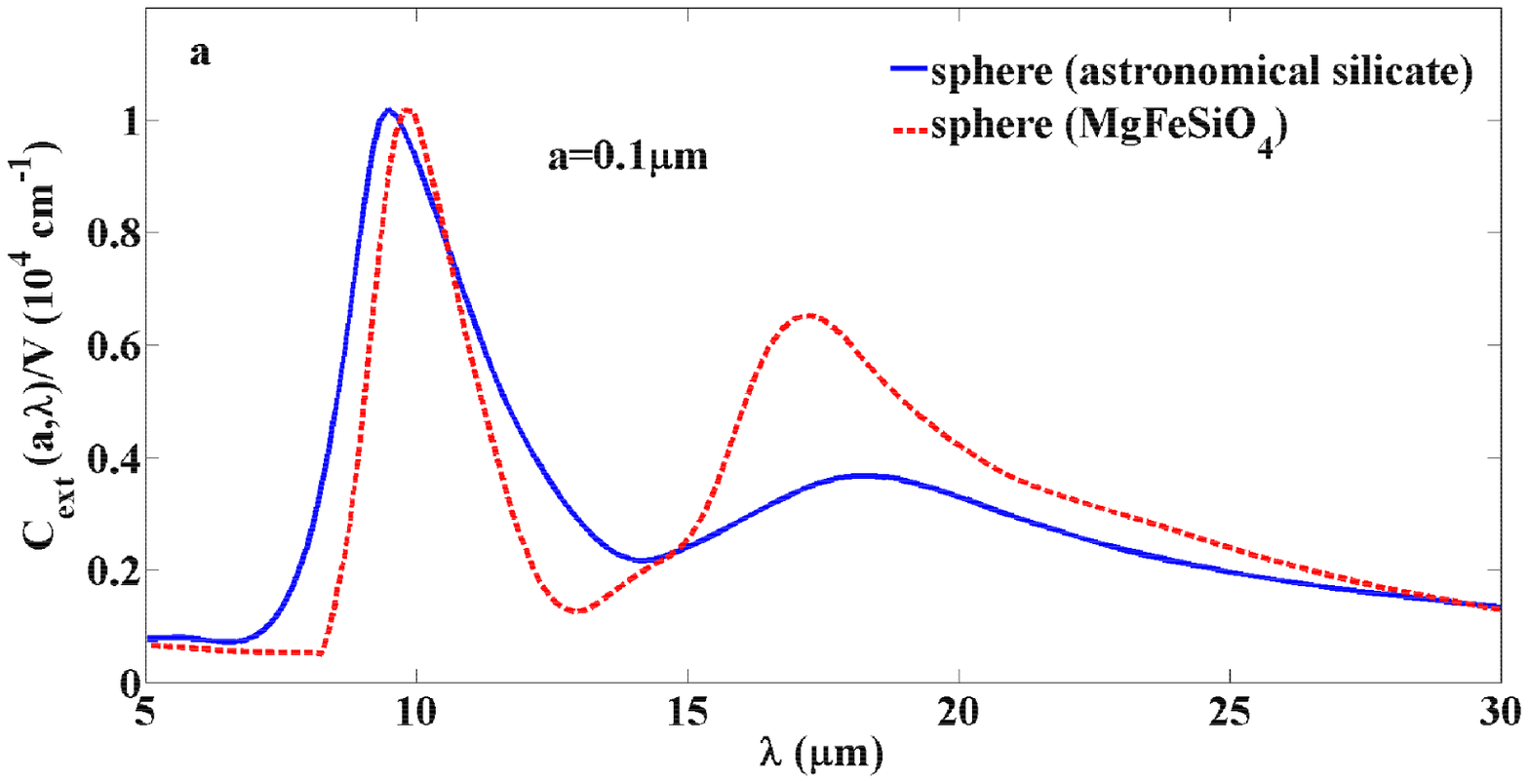}
\includegraphics[width=.8\textwidth]{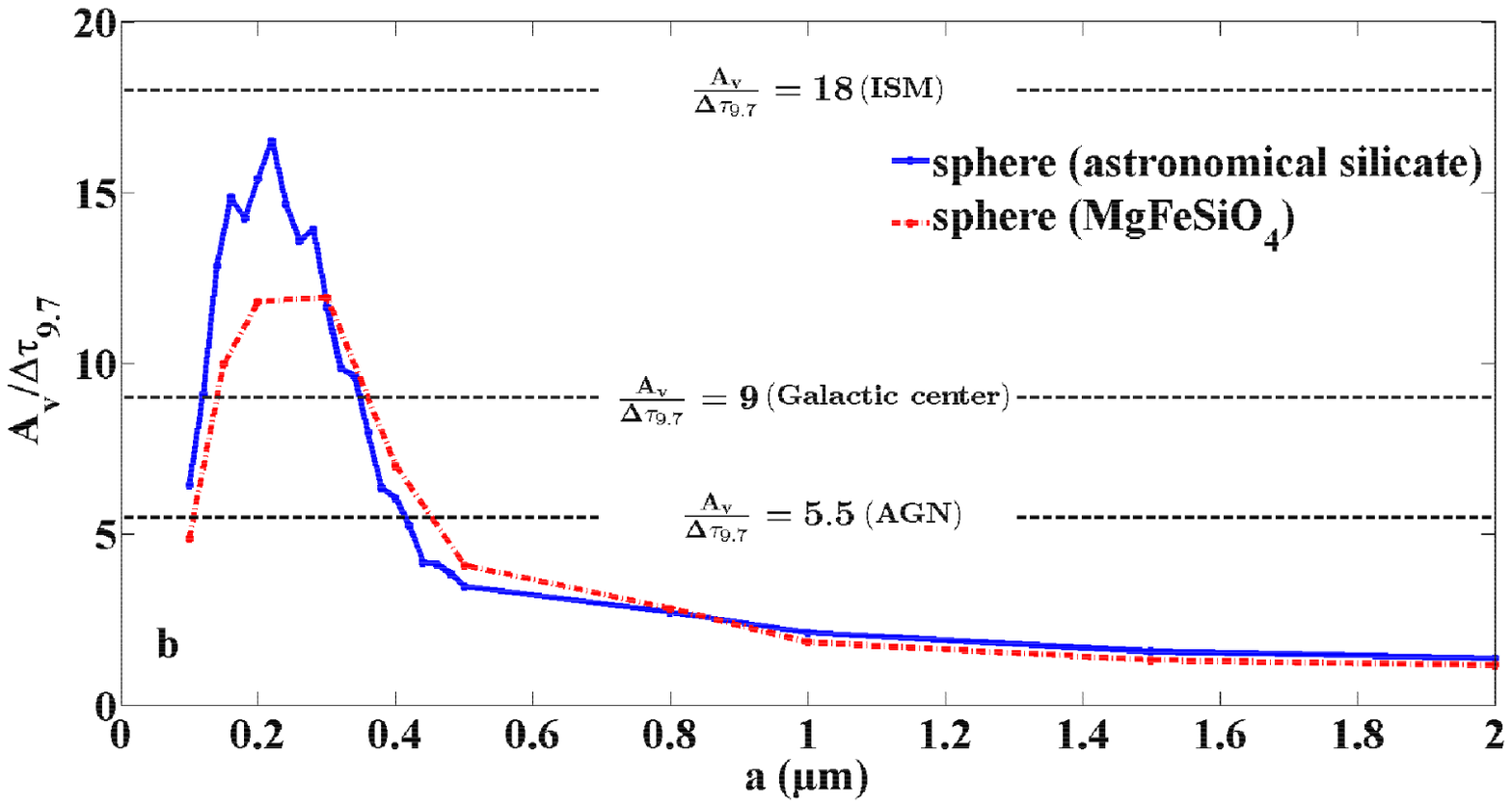}
\caption{\footnotesize
              \label{fig:composition}
              Upper panel (a):
              Comparison of the extinction profile
              of spherical amorphous olivine dust
              of radius $a=0.1\mum$ (red dotted line)
              with that of ``astronomical silicate''
              (blue solid line).
              Lower panel (b):
              Comparison of the variation of
              $\AV/\tausil$ with dust size $a$
              of spherical amorphous olivine dust
              (red dot-dashed line)
              with that of ``astronomical silicate''
              (blue solid line).
              }
\end{figure}

We have also investigated the effects
of silicate composition on $\AV/\tausil$
by considering the dielectric functions
of amorphous olivine MgFeSiO$_4$
experimentally measured by
Dorschner et al.\ (1995).
As shown in Figure~\ref{fig:composition}a,
compared to that of ``astronomical silicate'',
the 9.7$\mum$ Si--O feature of
spherical amorphous olivine of $a=0.1\mum$
peaks at a longer wavelength
and is considerably narrower,
also, the 18$\mum$ O--Si--O
feature is substantially stronger.
In Figure~\ref{fig:composition}b
we show the variation of $\AV/\tausil$
with the dust size $a$. Compared with
that of ``astronomical silicate'',
although the peak ratio of $\AV/\tausil$
of amorphous olivine is somewhat lower,
the overall profile of $\AV/\tausil$
of amorphous olivine resembles
that of ``astronomical silicate'',
particularly, for amorphous olivine,
one also requires the dust size to be
$>$\,0.4$\mum$ in order for
the calculated $\AV/\tausil$ to
agree with that of AGNs
(i.e., $\AV/\tausil\approx5.5$, Lyu et al.\ 2014).
This demonstrates that the exact silicate
composition does not considerably
affect $\AV/\tausil$.

\section{Discussion}\label{sec:discussion}
In \S\ref{sec:mod} we have shown that
the observed AGN ratio of
$\AV/\tausil\approx5.5$ (Lyu et al.\ 2014)
implies a dust size of $a\sim0.4\mum$,
substantially exceeding the characteristic size
of $\simali$0.1$\mum$ of interstellar grains.
This conclusion was derived from pure silicate dust.
As amorphous silicate and some sorts of
carbonaceous dust are the major dust species
of the diffuse ISM (e.g., see Mishra \& Li 2015,
Siebenmorgen et al.\ 2014),
one naturally expects both silicate dust and
carbon dust (e.g., graphite, amorphous carbon)
to be present in the dust torus around AGNs.
Indeed, the 3.4$\mum$ absorption feature,
commonly attributed to the C--H stretching
mode in saturated aliphatic hydrocarbon dust,
is seen in AGNs (Wright et al.\ 1996,
Imanishi et al.\ 1997). Mason et al.\ (2004)
argued that the 3.4$\mum$ absorption feature
at least in face-on Seyfert 2 galaxies arises in dust
local to the active nucleus rather than
in the diffuse ISM of the galaxy.
Since carbon dust contributes to $\AV$
but not to $\tausil$, with carbon dust included
one expects a higher ratio of $\AV/\tausil$
for a given dust size, and therefore, the observed
AGN ratio of $\AV/\tausil\approx5.5$ would require
the dust size to be even larger than that derived in
\S\ref{sec:mod} based on a pure silicate composition.

We now consider a mixture of
spherical amorphous carbon and
spherical ``astronomical silicate'' dust
with a mass ratio of $\Mcarb/\Msil$.
We calculate $\CextVcarb$ and $\CextVsil$,
the extinction cross sections of
amorphous carbon grains
and ``astronomical silicate'' grains
in the visual band, from Mie theory using
the dielectric functions of Rouleau \& Martin (1991)
for amorphous carbon and of Draine \& Lee (1984)
for ``astronomical silicate''.
Assuming that both the silicate dust component
and the carbon dust component have the same
size distribution, one derives $\AV/\tausil$ from
\begin{equation}
\frac{\AV}{\tausil} = \left(\frac{\AV}{\tausil}\right)_{\rm sil}
\times\left\{1+\frac{\rhosil}{\rhocarb}\cdot\frac{\Mcarb}{\Msil}
\cdot\frac{\CextVcarb}{\CextVsil}\right\} ~~,
\end{equation}
where $\left(\AV/\tausil\right)_{\rm sil}$ is
the ratio of the visual extinction
to the 9.7$\mum$ silicate optical depth
of the silicate dust component alone,
$\rhosil\approx3.5\g\cm^{-3}$ and
$\rhocarb\approx1.8\g\cm^{-3}$ are
respectively the mass densities of
amorphous silicate and amorphous carbon.

\begin{figure}[ht]
\centering
\includegraphics[width=.8\textwidth]{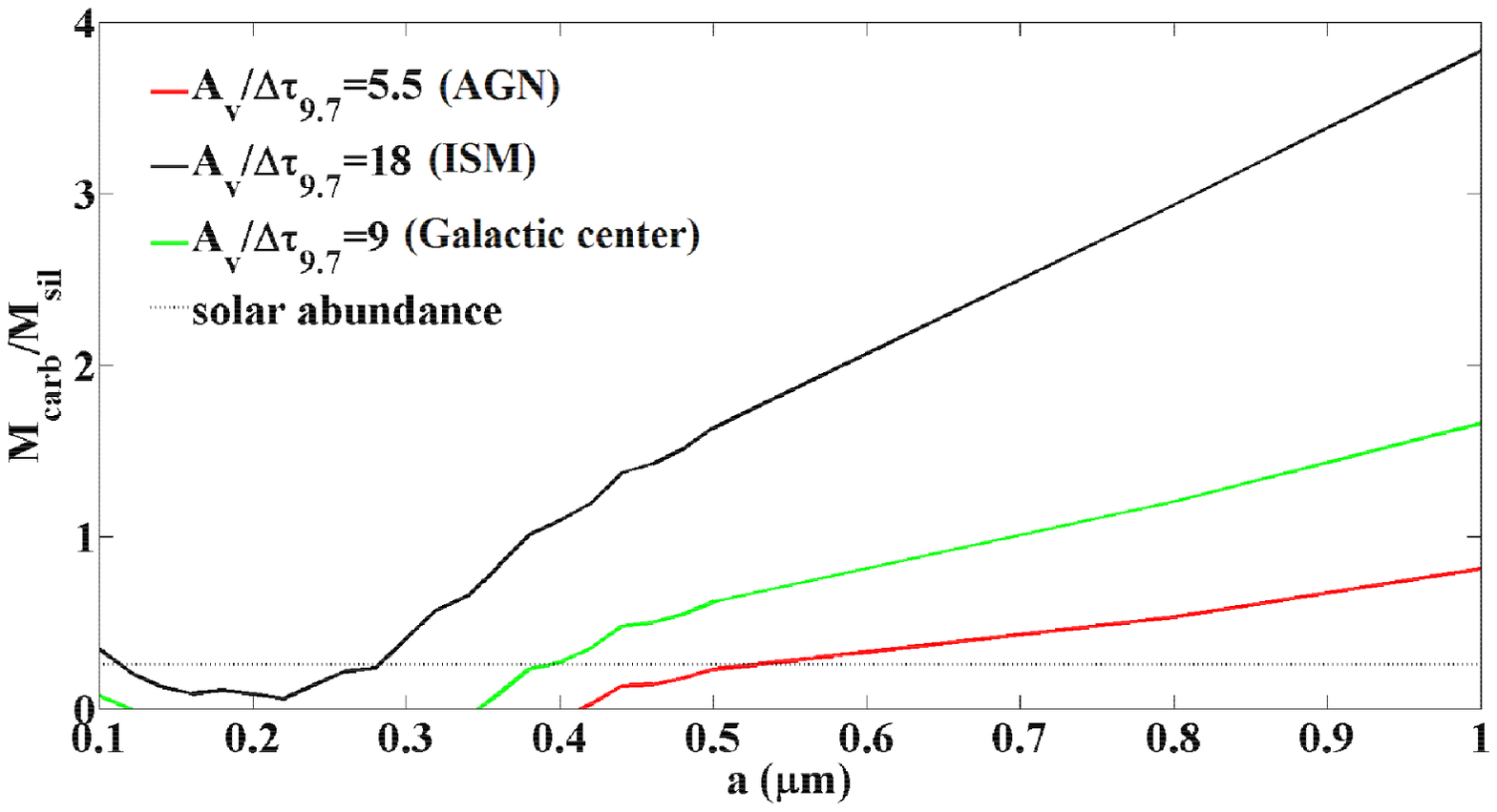}
\caption{\footnotesize
              \label{fig:carb}
              For a given carbon-to-silicate dust mass ratio
              $\Mcarb/\Msil$, the corresponding dust size
              required to account for the observed ratio of
              $\AV/\tausil\approx5.5$ of AGNs (red solid line),
              $\AV/\tausil\approx9$ of the Galactic center
              (green solid line), and
              $\AV/\tausil\approx18$ of the local ISM (black solid line).
              The gray dotted horizontal line plots
              the carbon-to-silicate dust mass ratio
              of $\Mcarb/\Msil\approx0.27$.
              }
\end{figure}

Apparently, the observed
$\left(\AV/\tausil\right)_{\rm obs}$ ratio
also provides information
about $\Mcarb/\Msil$,
the carbon dust to silicate dust mass ratio:
\begin{equation}
\frac{\Mcarb}{\Msil} =
\frac{\left(\AV/\tausil\right)_{\rm obs}/
\left(\AV/\tausil\right)_{\rm sil} -1}
{\left(\rhosil/\rhocarb\right)
\left\{\CextVcarb/\CextVsil\right\}} ~~.
\end{equation}
For a given dust size,  we show
in Figure~\ref{fig:carb}
the maximum allowable $\Mcarb/\Msil$
for accounting for the observed
$\left(\AV/\tausil\right)_{\rm obs}\approx5.5$
ratio of AGNs. As expected, a larger
$\Mcarb/\Msil$ ratio requires
the presence of larger grains.
With $\Mcarb/\Msil=0$, the observed ratio
of $\left(\AV/\tausil\right)_{\rm obs}\approx5.5$
implies $a\approx0.42\mum$.
Since it is unlikely for an AGN dust torus
to only have silicate dust, we conclude that
the dust size of $a\approx0.42\mum$
is a lower limit.

If the AGN torus has a solar abundance
for the dust-forming elements such as
C, O, Mg, Si, and Fe (Asplund et al.\ 2009),
one would obtain $\Mcarb/\Msil\approx0.27$ from
\begin{equation}\label{eq:Msil2MH}
\Mcarb/\Msil = \frac{\muc\left\{\cism-\cgas\right\}}
               {\mufe\fedust + \mumg\mgdust
                + \musi\sidust + 4\times \mu_{\rm O}\sidust} ~~,
\end{equation}
where $\cism$ is the interstellar C abundance (relative to H),
$\cgas$ is the gas-phase C abundance,
$\fedust$, $\mgdust$ and $\sidust$ are
respectively the abundances of Fe, Mg and Si
locked up in silicate dust of
a stoichiometric composition of
Mg$_{\rm 2x}$Fe$_{\rm 2(1-x)}$SiO$_4$
(i.e., each silicon atom
corresponds to four oxygen atoms),
and $\muc$, $\mufe$, $\mumg$,
$\musi$ and $\mu_{\rm O}$ are respectively
the atomic weights of C, Fe, Mg, Si and O.
We adopt the solar abundance of
$\csun\approx269\pm31$
parts per million (ppm),
$\fesun\approx31.6\pm2.9\ppm$,
$\mgsun\approx39.8\pm3.7\ppm$, and
$\sisun\approx32.4\pm2.2\ppm$
(Asplund et al.\ 2009),
and assume all Fe, Mg and Si elements
are depleted in silicate dust
(i.e., $\fedust\approx\fesun$,
$\mgdust\approx\mgsun$,
$\sidust\approx\sisun$).
For C, subtracting the gas-phase abundance
of $\cgas\approx140\ppm$ (Cardelli et al.\ 1996),
we are left with
$\cdust=\cism-\cgas\approx\csun-\cgas\approx129\ppm$
for the carbon dust component.
%
%
%
With $\Mcarb/\Msil\approx0.27$,
one infers the dust size to be
$a\approx0.54\mum$ in order to
account for the observed AGN ratio
of $\AV/\tausil\approx5.5$
(see Figure~\ref{fig:carb}).
Figure~\ref{fig:carb} also shows that
a larger $\Mcarb/\Msil$ ratio
(e.g., due to a lower gas-phase abundance
[e.g., see Sofia et al.\ 2011]) would
require a larger grain size.

\begin{figure*}
\centering
\includegraphics[width=.8\textwidth]{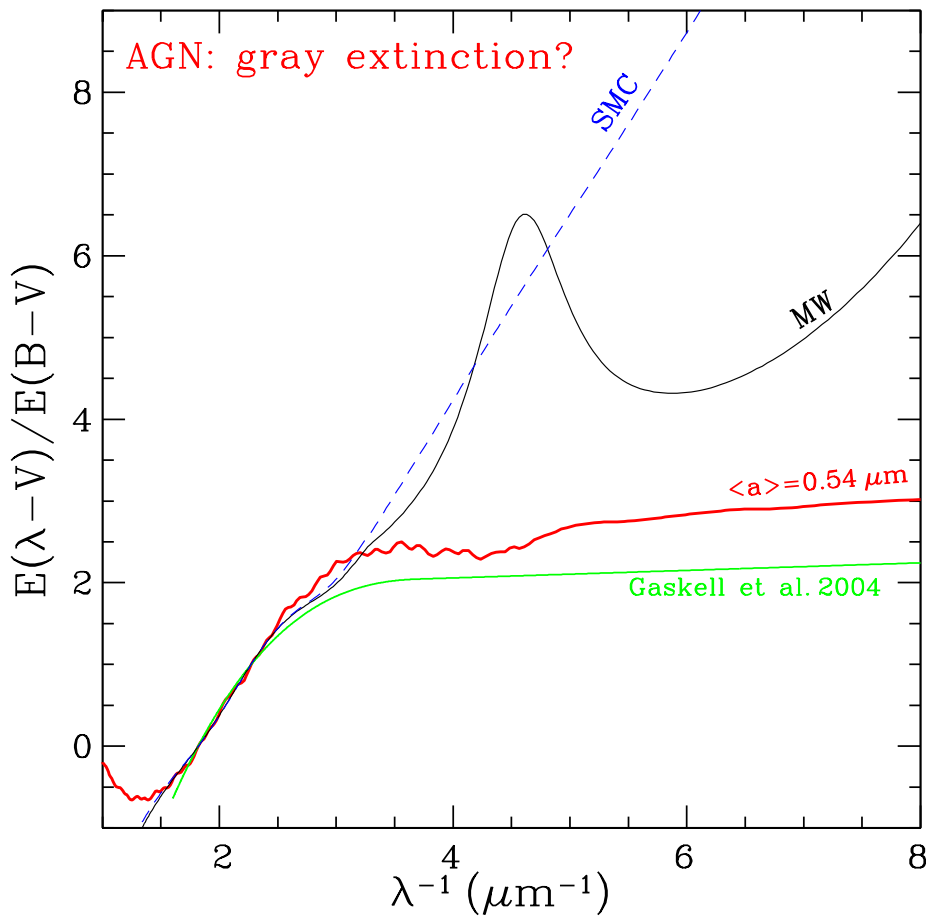}
\caption{\footnotesize
         \label{fig:extcurv}
         Comparison of the extinction curves
         of the Milky Way (black solid line)
         and the SMC (blue dashed line)
         with that of Gaskell et al.\ (2004)
         derived from composite quasar spectra
         (green solid line) and that calculated
         from spherical ``astronomical silicate'' grains
         and spherical amorphous carbon grains
         of area-weighted mean radii
         $\langle a\rangle=0.54\mum$
         and of a mass mixing ratio
         $\Mcarb/\Msil=0.27$ (red solid line).
         }
\end{figure*}

We now calculate the extinction as a function of
$\lambda^{-1}$ expected from a mixture of
spherical ``astronomical silicate''
and spherical amorphous carbon
of area-weighted mean radii
$\langle a\rangle = 0.54\mum$
and of a mass ratio of $\Mcarb/\Msil=0.27$.
We represent the extinction
by $E(\lambda-V)/E(B-V)$,
where $E(\lambda-V)\equiv A_\lambda - A_V$,
$E(B-V)\equiv A_B - A_V$,
and $A_B$ is the extinction at the $B$ band
(i.e., $\lambda_B\approx4400\Angstrom$).
To avoid the resonant structures expected
to be pronounced in the extinction profiles
of grains of a single size
(see Bohren \& Huffman 1983),
we consider a MRN-type size distribution of
$dn/da \propto a^{-3.5}$ (Mathis et al.\ 1977)
in the size range of $0.2\simlt a\simlt1.5\mum$.
This gives an area-weighted mean size
of $\langle a\rangle\approx0.54\mum$.
As shown in Figure~\ref{fig:extcurv},
the extinction curve predicted from
such a mixture is flat or gray
at $\lambda^{-1}>3\mum^{-1}$
(i.e., the extinction varies little
with $\lambda^{-1}$)
and closely agrees with that of
Gaskell et al.\ (2004) who derived an
AGN extinction curve based on the composite spectra
of 72 radio quasars and 1018 radio-quiet AGNs.
We note that Czerny et al.\ (2004) also constructed
a featureless flat extinction curve for quasars
based on the blue and red composite quasar spectra of
Richards et al.\ (2003) obtained
from the Sloan Digital Sky Survey (SDSS).
The extinction curves of both the Milky Way
and the Magellanic Clouds differ substantially
from our model extinction curve as well as that of
Gaskell et al.\ (2004) and Czerny et al.\ (2004)
in that, the Galactic extinction curve shows
a prominent extinction bump at 2175$\Angstrom$
and a steep far-UV rise which is believed to arise
from small carbon grains (e.g., see Mishra \& Li 2015).
In contrast, the extinction curve of
the SMC lacks the 2175$\Angstrom$ bump
and displays an even steeper far-UV rise
than that of the Milky Way.

Observationally, the predominant presence of
large grains in AGN torus has also been indirectly
inferred from the substantially reduced
dust reddening- and extinction-to-gas ratios of AGNs:
Maiolino et al.\ (2001a)  determined for 19 AGNs
the amount of reddening $E(B-V)$, $\AV$,
and $N_{\rm H}$, the hydrogen column densities.
They found that for most (16 of 19) objects
both $E(B-V)/\NH$ and $\AV/\NH$ are significantly
lower than that of the Galactic diffuse ISM
by a factor ranging from a few to $\simali$100
(but also see Weingartner \& Murray 2002).
Maiolino et al.\ (2001b) ascribed the reduced
$E(B-V)/\NH$ and $A_V/\NH$ ratios of AGNs
(often with a solar or higher metallicity)
to grain growth through
coagulation in the dense circumnuclear region which
results in a dust size distribution biased in favour
of large grains and therefore a flat extinction curve.
We note that, this could also be caused by
the preferential destruction of small grains
in the hostile torus which is exposed to
the energetic X-ray and far-UV photons
from the central engine.

It is interesting to note that, as shown in
Figures~\ref{fig:AV2Sil},\,\ref{fig:shape}b,\,\ref{fig:composition}b,
silicate dust alone is not capable of explaining
the observed ratio of $\AV/\tausil\approx18$
of the local solar neighborhood diffuse ISM
(Roche \& Aitken 1984).
This is true for both spherical and nonspherical
grains and for both ``astronomical silicate''
and amorphous olivine compositions.
With carbon dust included,
$\AV/\tausil\approx18$
is now achievable provided $\Mcarb/\Msil\simgt0.15$.
For a solar abundance of $\Mcarb/\Msil\approx0.27$,
a mixture of silicate and carbon grains
of sizes $a\approx0.11\mum$
or $a\approx0.28\mum$ result
in $\AV/\tausil\approx18$
(see Figure~\ref{fig:carb}).
We disregard the size of $a\approx0.28\mum$
since the interstellar extinction curve constrains
the mean dust size to be $\simali$0.1$\mum$.\footnote{%
   There may exist a population of
   very large grains in the ISM
   which substantially exceed 0.1$\mum$
   (e.g., see Gr\"un et al.\ 1994, Block et al.\ 1994,
    Witt et al.\ 1994, Taylor et al.\ 1996,
    Witt et al.\ 2001, Wang et al.\ 2014,
    Westphal et al.\ 2014, Sterken et al.\ 2015,
    Wang et al.\ 2015a,b, Krelowski 2017).
    However, these grains are gray and do not
    show up in the UV/optical extinction curve.
    Also, they only account for $\simali$15\%
    of the total dust mass in the ISM
    (e.g., see Wang et al.\ 2015a,b).
    }

Within the Galaxy,
not only the 9.7$\mum$ silicate extinction profile
but also the $\AV/\tausil$ ratio
vary among different sightlines.
With $\AV/\tausil\approx9$ (Roche \& Aitken 1985),
the optical-to-silicate extinction ratio
of the line of sight toward the Galactic center
differs from that of the local ISM
by a factor of $\simali$2.
Roche \& Aitken (1985) suggested that
the reduction of $A_V/\tausil$
in the Galactic center may be caused
by the fact that there are fewer carbon stars
in the central regions of the Galaxy and therefore,
the production of carbon-rich dust may be
substantially reduced compared with
the outer Galactic disk.
However, Gao et al.\ (2010) argued against
this hypothesis based on the fact that the 3.4$\mum$
C--H feature of aliphatic hydrocarbon dust
also exhibits a similar behavior:
$\AV/\tauahc\approx146$ of the Galactic center
also differs from that of the local ISM of
$\AV/\tauahc\approx274$
by a factor of $\simali$2,
where $\tauahc$ is the optical depth
of the 3.4$\mum$ aliphatic C--H
absorption feature.
If the hypothesis of Roche \& Aitken (1985)
was correct, one would expect a larger $\AV/\tauahc$ ratio
in the Galactic center than that of the local ISM.
We note that, along the lines of sight
toward the Galactic center,
there are dense molecular clouds.\footnote{%
  The sightline toward the Galactic center source
  $\sgrA$ suffers about $\simali$30$\magni$ of
  visual extinction (e.g. see McFadzean et al.\ 1989),
  to which molecular clouds may contribute as much as
  $\simali$10$\magni$ (Whittet et al.\ 1997).
  }
In cold, dense molecular clouds, interstellar dust
is expected to grow through coagulation
(as well as accreting an ice mantle).
Indeed, as shown in Figure~\ref{fig:carb},
grains of $a\simgt0.35\mum$ are able to
account for
the observed ratio of $\AV/\tausil\approx9$.
%
For a solar abundance of $\Mcarb/\Msil\approx0.27$,
a mixture of silicate and carbon grains
of sizes $a\approx0.40\mum$
lead to $\AV/\tausil\approx9$
(see Figure~\ref{fig:carb}).
%

\begin{figure}[ht]
\centering
\includegraphics[width=.8\textwidth]{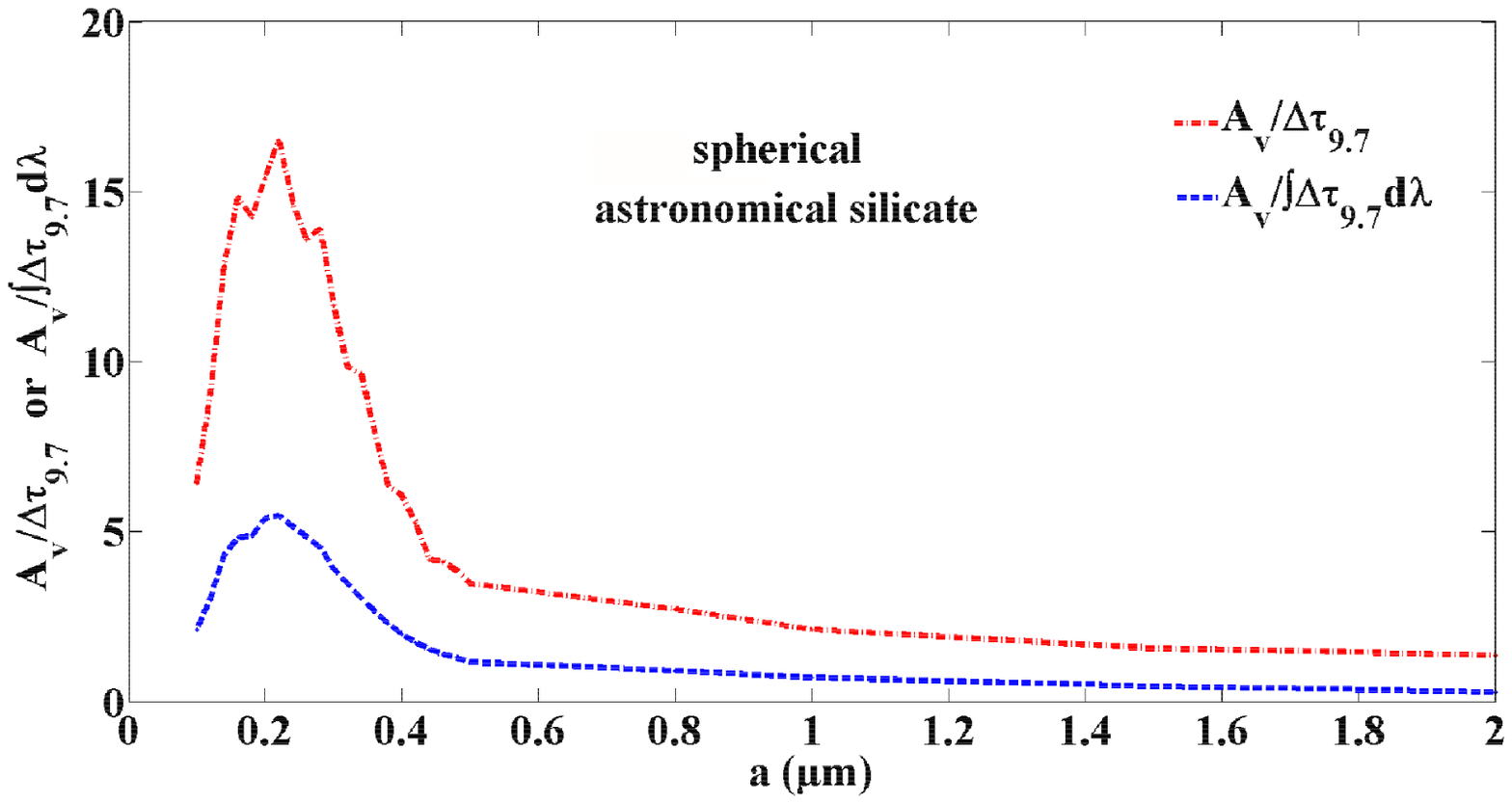}
\caption{\footnotesize
               \label{fig:integral}
               Comparison of $\AV/\int\tausil d\lambda$
               with $\AV/\tausil$.
               }
\end{figure}

Finally, we argue that $\tausil$ may not
be an accurate measure of the silicate
absorption. Observationally, it is well
recognized that the FWHM ($\gamma$)
of the 9.7$\mum$ silicate absorption feature
varies from one line of sight to another
[e.g., $\gamma\simali$2.6$\mum$ for
the local diffuse ISM along the line of sight
toward Cyg OB2 \#12 (Whittet et al.\ 1997),
$\gamma\simali$1.8$\mum$ for the sightline toward
the Galactic center object Sgr A$^{\star}$
(Kemper et al.\ 2004)].
Theoretically, the FWHM of the 9.7$\mum$
silicate feature computed from Mie theory
increases with the dust size
(see Figure~\ref{fig:Cextsil}).
We suggest that a more accurate measure
of the silicate absorption would be
$\int\Delta\tau_{9.7}d\lambda$,
the integral of the continuum-subtracted
9.7$\mum$ silicate absorption feature.
Similarly, we obtain the visual extinction $\AV$
also by integrating $A_\lambda$ over
the Johnson $V$-band filter response function.
In Figure~\ref{fig:integral} we compare
$\AV/\int\tausil d\lambda$ with $\AV/\tausil$.
As expected, $\AV/\int\tausil d\lambda$
is not simply scaled from $\AV/\tausil$.
We argue that, in future observational studies,
it is more meaningful to determine
$\int\tausil d\lambda$ than $\tausil$.

\section{Summary}\label{sec:summary}
We have explored the possible
diagnostic power of $\AV/\tausil$,
the ratio of the optical extinction
in the visual band to the optical depth
of the 9.7$\mum$ silicate absorption feature,
in gaining insight into the size of the dust
in AGNs, the local ISM, and the Galactic center.
It is shown that,
with $\AV/\tausil\approx5.5$,
the dust in AGN torus could exceed
$\simali$0.4$\mum$ in radius,
substantially larger than that of
the Galactic diffuse ISM
for which $\AV/\tausil\approx18$, and
indicating a flat extinction curve for AGNs.
Similarly, the dust associated with
the Galactic center sightlines
for which $\AV/\tausil\approx9$
could also be considerably
larger than that of the Galactic diffuse ISM,
suggesting the coagulational growth
of the dust in the molecular-cloud components
along the lines of sight toward the Galactic center.

\acknowledgments{
We thank J.~Gao, L.~Hao, J.W.~Lyu,
S.~Wang, Y.X.~Xie
and the anonymous referee
for stimulating discussions
and very helpful suggestions.
This work is supported by NSFC through
Projects 11173007, 11373015, 11533002,
and 973 Program 2014CB845702.
AL is supported in part by
NSF AST-1311804 and NASA NNX14AF68G.
}


\end{document}